\documentclass[aps,prl,twocolumn,showpacs,superscriptaddress,groupedaddress]{revtex4}  % for review and submission
\usepackage{graphicx}  % needed for figures
\usepackage{dcolumn}   % needed for some tables
\usepackage{bm}        % for math
\usepackage{amssymb}   % for math
\usepackage{amsmath}
\usepackage{subfigure}
\usepackage{color}

\begin{document}

% The following information is for internal review, please remove them for submission
%\centerline{\em INTERNAL DOCUMENT -- NOT FOR PUBLIC DISTRIBUTION}

% the following line is for submission, including submission to the arXiv!!

\title{Coarsening dynamics of three dimensional levitated foams: from wet to dry}

\author{N. Isert} \affiliation{Fachbereich Physik, University of Konstanz,
Universit\"atsstrasse 10, 78457 Konstanz, Germany}
\author{G.~Maret} \affiliation{Fachbereich Physik, University of Konstanz,
Universit\"atsstrasse 10, 78457 Konstanz, Germany}
\author{C.M.~Aegerter} \affiliation{Physik-Institut, University of Z{\"u}rich, Winterthurerstrasse 190, 8057 Z\"urich, Switzerland}

\date{\today}

\begin{abstract}
We study diamagnetically levitated foams with widely different
liquid fractions. Due to the levitation, drainage is effectively
suppressed and the dynamics is driven by the coarsening of the
foam bubbles. For dry foams, the bubble size increases as the
square root of foam age, as expected from a generalized von
Neumann law. At higher liquid content, the behavior changes to
that of Ostwald ripening where the bubbles grow with the $1/3$
power of the age. Using Diffusing Wave Spectroscopy we study the
local dynamics in the different regimes and find diffusive
behavior for dry foams and kinetic behavior for wet foams.
\end{abstract}

\pacs{83.80.Iz, 05.70.Ln, 83.85.Ei, 64.75.Cd}
\maketitle

% sections are not used for PRL papers

Foams are models of soft matter consisting of gas bubbles enclosed
in a liquid, which have solid properties due to
surface tension of the bubbles and their inside pressure \cite{foam}.
Depending on the amount of liquid in the foam, bubbles can either be in contact or separated. In the former most common
case of "dry" or "almost dry" foams , bubbles are substantially deformed and the thin flat liquid films in the contact area between them provide substantial mechanical stability to the foam \cite{foam}. "Wet" foams are less stable because the rather homogeneous mixture of separated bubbles is rapidly destroyed by flows due to buoyancy pushing bubbles upwards \cite{foam}.
In "wet" and "almost dry" foams the dynamics is thus
driven by the drainage of liquid between the bubbles due to
gravity \cite{drainage}. However, additional dynamics occurs even without drainage because of gas exchange between bubbles
\cite{twodcoarsening} due to the Laplace pressure  $\Delta p = 2\sigma/r$ with $r$ being the bubble´s radius of curvature and $\sigma$ the liquid-gas surface tension \cite{laplace}. As a consequence, the average bubble size
increases with time because of the lower gas pressure in the
larger bubbles. This process, known as coarsening, has been described by von Neumann \cite{neumann}
and has been experimentally observed in two dimensional foams
\cite{twodcoarsening} as well as for very dry foams
\cite{mricoarsening}.

Here, we study three dimensional foams,
which are levitated by a strong magnetic field gradient
\cite{braunbeck,geim}. Due to the diamagnetism of water, it is
possible to effectively suppress the buoyancy of the gas bubbles
and thus stabilize the foam against drainage even at high liquid
content. With this simple trick it becomes possible to study the coarsening dynamics in 3D foams without chemical stabilizers for dry as well as wet foams over many hours in the laboratory. Without levitation such foams would decay within minutes on earth \cite{raphael} which is why major efforts are under way to investigate them at conditions of steady microgravity at the ISS \cite{microgravity}.

In dry foams the exchange of gas between bubbles
takes place directly through the thin liquid films separating the
bubbles. Because the rate of exchange is governed by the
(Laplace-)pressure $\Delta p$, the growth rate of a bubble,
i.e. the current density of gas exchange, is proportional to the
inverse of its radius, i.e. $j = \frac{dV}{A dt} \propto dr/dt
\propto 1/r$. Here, $A$ is the contact area of a bubble
which is of order $r^2$ and $V$ its volume.
Given this dynamics,
one obtains that the average size
of bubbles will increase with time as $\langle r \rangle \propto
t^{1/2}$. This can also be derived more strictly, as for instance
done in \cite{generalneumann}. When the bubbles are no longer in
contact, it can be surmised that the mechanism of gas exchange
between bubbles will have to change. In fact, the exchange of gas
will now have to be achieved via diffusion in the liquid and the
difference of the gas pressure in the bubbles to the saturation
pressure in the liquid. Here, the diffusive current density $j
\propto dr/dt$ will be determined by the gradient of the
concentration, i.e. pressure difference. This means that we obtain
$j \propto dr/dt \propto dp/dr$. Again using the fact that the
pressure inside the bubbles is given by Laplace's law, we obtain
$dr/dt \propto 1/r^2$ and hence a growth of the form $\langle r
\rangle \propto t^{1/3}$. Again, this can be derived studying the
detailed dynamics \cite{ostwald}. This qualitatively different
type of coarsening is also known as Ostwald ripening and is for
instance observed in the dynamics of the growth of inclusions in
solids \cite{solidcoarsening}. Since the nature of the dynamics
changes when the bubbles are no longer in contact, the boundary
between the two regimes is expected to be at a liquid fraction of
$\sim 30 \%$, which corresponds to the inverse density of closely
packed spheres \cite{closepacking}. Since levitated foams can be
created with a varying amount of liquid and relative stability to
drainage, these predictions can thus be tested experimentally with
our setup.

\begin{figure}
    \includegraphics[width=\linewidth]{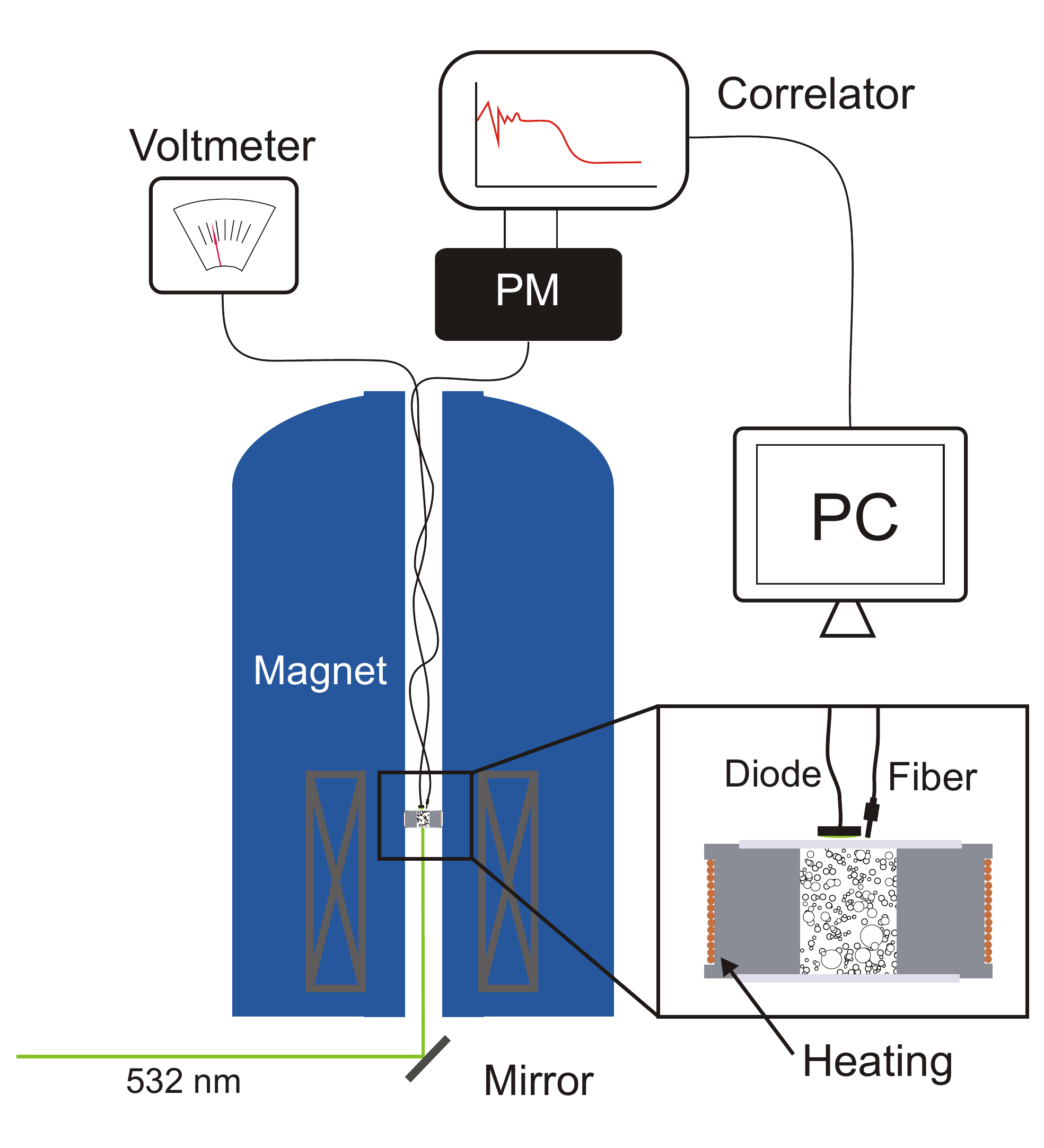}
    \caption{Schematic illustration of the levitation setup. The
    foam sample cell, shown in detail in the inset, is placed inside
    the free bore of a 18 T cryomagnet, with optical access from
    above and below. From below, a CW laser is incident on the sample
    cell, whose transmission through the sample as a function of time
    is measured from above by a photodiode and an optical fibre
    connected to a photomultiplier and a correlator. For the determination
    of the integrated transmitted intensity, the photodiode is used,
    whereas the autocorrelation function from the correlator gives
    information about the dynamics discussed below. The inset
    shows details of the sample cell. Windows on top and bottom allow for
    optical access and a small coil in the side of the sample cell
    is used to heat the sample in the bore of the magnet, such that
    it is always kept at temperatures above 30$^\circ$C. \label{setup} }
\end{figure}

In order to assess the growth of the foam bubbles with age inside
the three dimensional foam, we use multiply scattered light
transmitted through the foam \cite{Ohm}. The transport mean free
path of light in the turbid medium of the foam has been found to
be proportional to the bubble size \cite{durian,cohen}. This holds
both for dry foams, where scattering mainly takes place at the
thin liquid films between bubbles as well as for liquid foams,
where the scattering centers are the bubbles themselves. In
addition, the transmitted diffuse light can be used to obtain
information on the averaged local dynamics of the foam via
Diffusing Wave Spectroscopy (DWS) \cite{DWS}. Here, the
auto-correlation function of the multiply scattered light field
directly provides the time dependence of the phase shift incurred
by the dynamics of the scattering particles. Thus it is possible
to determine whether the dynamics of the scatterers is diffusive
or ballistic and from the respective time scales \cite{DWS2}, the
size of the dynamic particles can be obtained.

The foams used in the experiments consist of water, sodium dodecyl
sulfate (SDS) as a surfactant and N$_2$ gas. The water-SDS mixture
and the gas are put in two separate syringes, which are connected
through a thin tube \cite{SDS}. The water-SDS mixture is then
transferred to the gas-filled syringe and the resulting mixture is
transferred through the thin tube several times in order to
achieve turbulent mixing. In this way an irregular foam with a
given mean pressure in the bubbles determined by the pressure on
the syringes is obtained. The liquid content in the foam is
determined by the volume-ratio of water-SDS mixture and gas in the
initial state within the two syringes \cite{SDS}. Due to the
compression of the gas by the syringes, the effective liquid
fraction of the foam will be somewhat higher than the initial
composition given by the content in the two syringes. An initial
liquid fraction of 25$\%$ corresponds roughly to an effective
liquid fraction of 30$\%$. The foam thus created is then
transferred to a sample-cell of a diameter of 1.7 cm and a height
of 1.2 cm, which is placed inside the room temperature bore of a
superconducting magnet capable of applying a field of 18 T. Due to
the insufficient thermal insulation of the bore and the high
freezing point of a water-SDS mixture, a small heating coil is
added around the sample cell in order to keep the foam at constant
temperature all times (see Fig. \ref{setup}). The magnet is a
superconducting solenoid at the end of which the field shows a
substantial gradient \cite{corinna}. This means that there is a
significant upward force on the diamagnetic water-SDS mixture,
given by $f = \chi B\partial B/\partial z$, where $f$ is the force
density and $\chi$ is the diamagnetic susceptibility. At a
specific point, when $B\partial B/\partial z = \rho g/\chi$, this
force exactly compensates the gravitational force
\cite{braunbeck}. Due to the field distribution,
this will lead to a stable levitation at this point \cite{geim},
where the levitation will be homogeneous to one part in a thousand
within a volume of 1 cm$^3$, thus for almost the whole foam
sample. Eliminating drainage this way opens up the possibility of observing the
coarsening dynamics over extraordinary long times.

\begin{figure}
    \includegraphics[width=\linewidth]{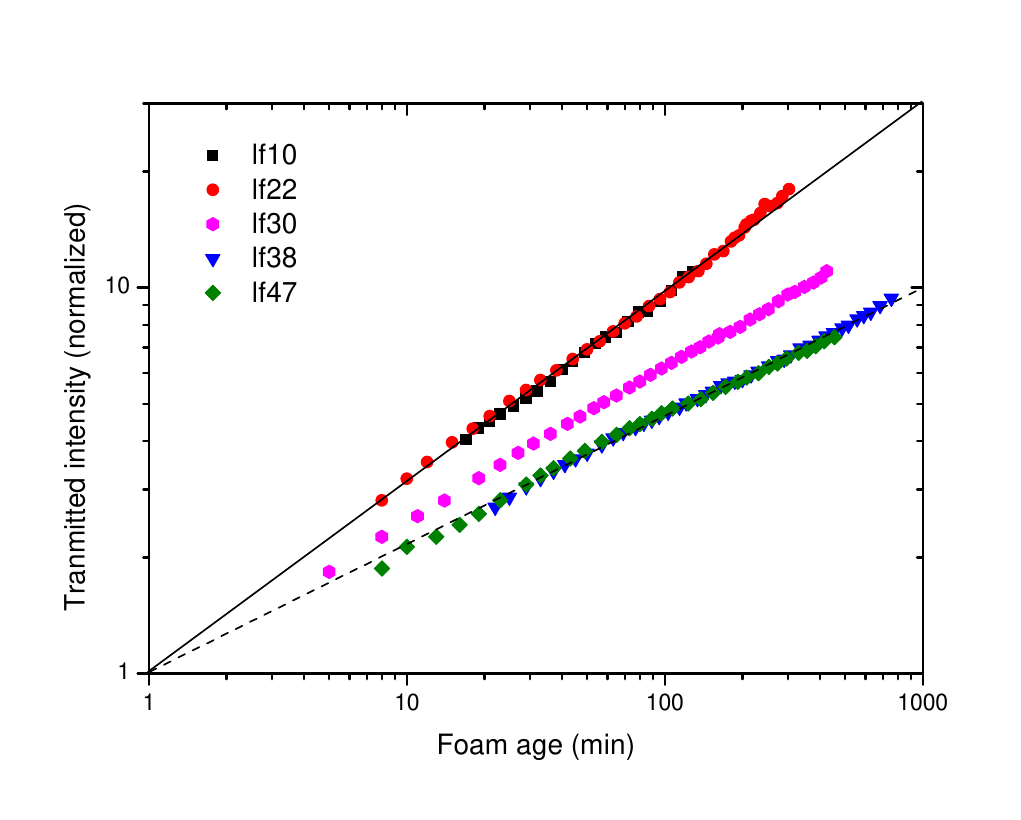}
    \caption{The overall size evolution of the bubbles of a levitated
    foam are determined via the mean free path, $l^*$, as a function
    of foam age. In the limit of small absorption, the transmitted
    intensity, which is plotted here is directly proportional to $l^*$.
    Moreover, it has been shown that $l^*$ in a foam is proportional to
    the bubble size \cite{durian2}, such that a temporal evolution of
    the transmitted intensity can be used to study the bulk coarsening
    dynamics directly. Here we show five examples of foams at different
    liquid fractions, showing different scaling behavior of $l^*$. As
    can be seen in foams with liquid content above 30$\%$ the bubbles
    increase in size more slowly, whereas at lower liquid fraction
    they increase faster. Above and below a narrow transition region,
    the behavior is however independent of liquid fraction. The slope
    in this double logarithmic plot directly gives the scaling exponent
    of the coarsening dynamics. Straight lines with a slope of 1/2
    (solid) and 1/3 (dashed) are added for comparison. \label{lstar} }
\end{figure}

In order to study the dynamics, the foam is illuminated with a
Coherent Verdi solid state CW laser at a wavelength of 532 nm and
a power of 100 mW and the transmitted light is detected either
with a photodiode or a glass fiber leading to a photomultiplier
and a correlator card for DWS measurements \cite{DWS3}. The
schematic setup is again shown in Fig. \ref{setup}. Here, the
auto-correlation function of fluctuating transmitted intensity is
determined. These fluctuations are due to small movements of the
scatterers (i.e. the foam bubbles), which lead to a change in the
interference pattern of different multiple scattering paths
through the sample \cite{DWS}. This means that the correlation
time gives a measure of the scatterer movement via the mean square
fluctuations in the phase of the light induced by these movements.

In the multiple scattering regime, the transmitted intensity is
given by Ohm's law, i.e. $T \propto l^*/L$ \cite{Ohm}, where $L$
is the thickness of the sample, $l^*$ is the transport mean free
path. Thus by determining the average transmitted intensity, we
can directly obtain a measure of the change of the mean free path
$l^*$ with foam age, since both the sample thickness and the
incident intensity are fixed. The mean free path of light in the
sample has been shown before to give a determination of the bubble
size with $l^* \propto r$ \cite{durian2}. In figure \ref{lstar},
this dependence of $l^*$ with foam age is shown for a set of foams
with different liquid content. As can be seen in this double
logarithmic plot, all foams show a scaling behavior with a power
law increase of bubble size with age. For dry foams, this increase
is faster with an exponent close to 1/2, whereas for wet foams it
is slower with an exponent close to 1/3. These exponents are the
asymptotic dynamics of the theoretical predictions for foam
dynamics in the dry and wet case respectively
\cite{neumann,ostwald}. In the transition region, there is an
intermediate behavior, where two distinct regimes can be seen
corresponding to the two different dynamics.

\begin{figure}
    \includegraphics[width=\linewidth]{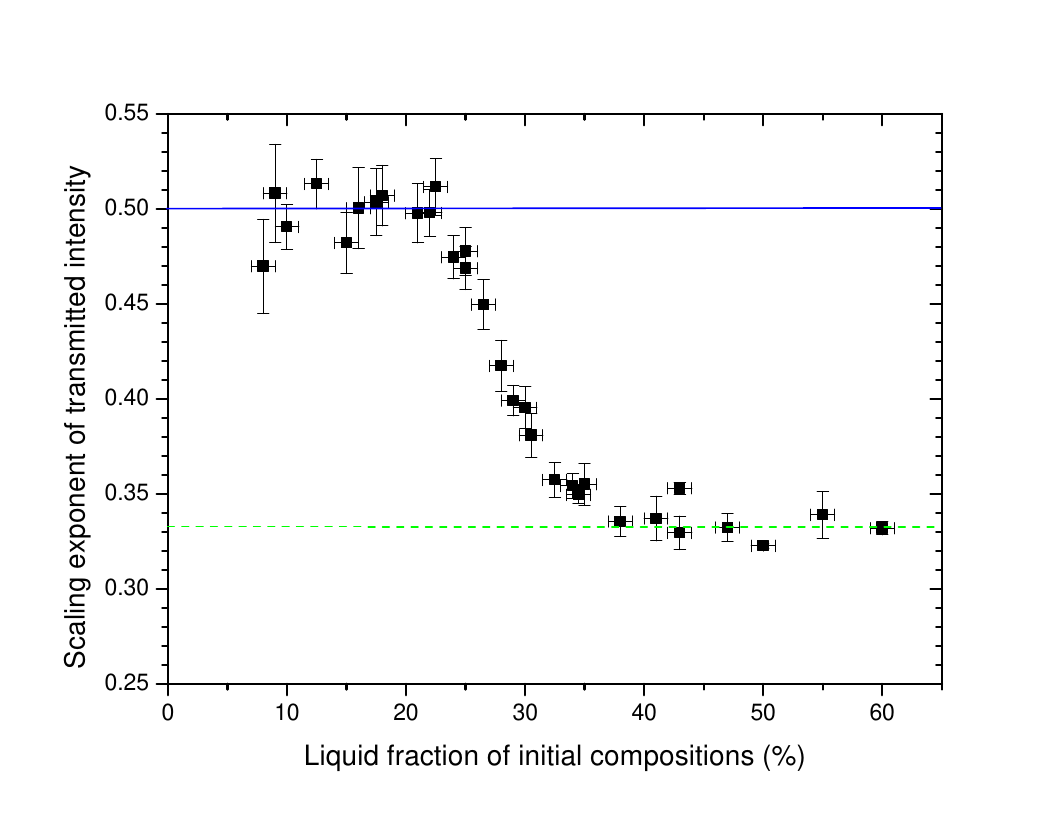}
    \caption{Data similar to that shown in Fig. \ref{lstar} but for
    a bigger variety of liquid fractions is summarized via the value
    of the scaling exponent as a function of the liquid fraction. At
    low liquid fractions, the exponent is in good
    accord with the von Neumann expectation of 1/2, whereas for high
    liquid fractions it is consistent with the expectation of Ostwald
    ripening of 1/3. The transition between these two regimes is
    rather narrow at liquid fractions between 25 and 30 $\%$, as is
    expected for the point where separation of bubbles begins to occur. \label{exponent} }
\end{figure}

These results are summarized in Fig. \ref{exponent}, where the
exponents fitted for all experiments within a large range of
liquid fractions is plotted. It can be clearly seen that below a
liquid fraction of 25$\%$, the exponents are all compatible with
1/2, whereas above an initial liquid fraction of 30$\%$, they are
all compatible with 1/3. Thus there is a clear transition in the
coarsening behavior of the foams at a liquid fraction
corresponding to the close packing of spheres. Both the transition
as well as the values of the exponents are predicted by theory
\cite{neumann,ostwald}.

\begin{figure}
    \includegraphics[width=0.83\linewidth]{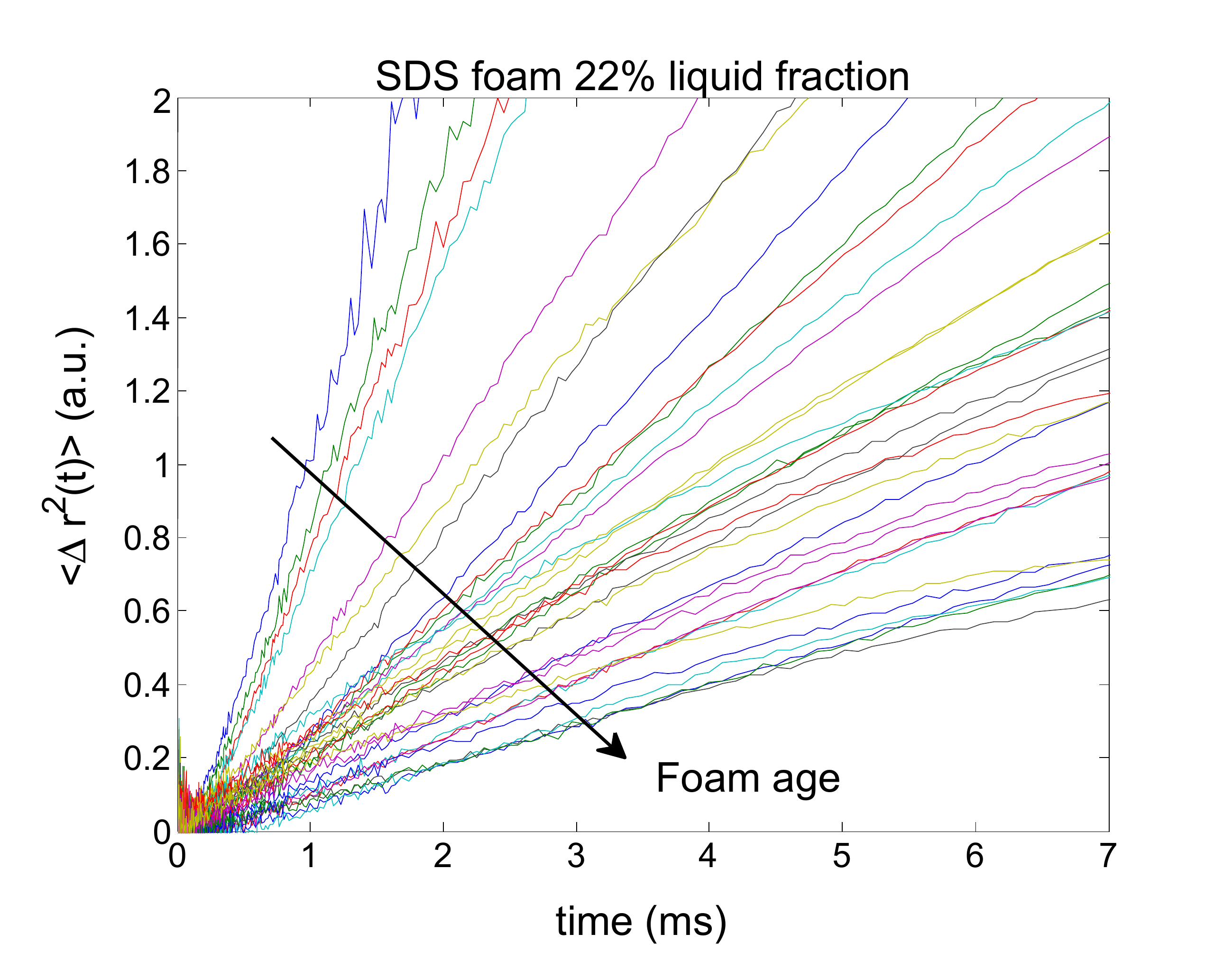}
    \includegraphics[width=0.83\linewidth]{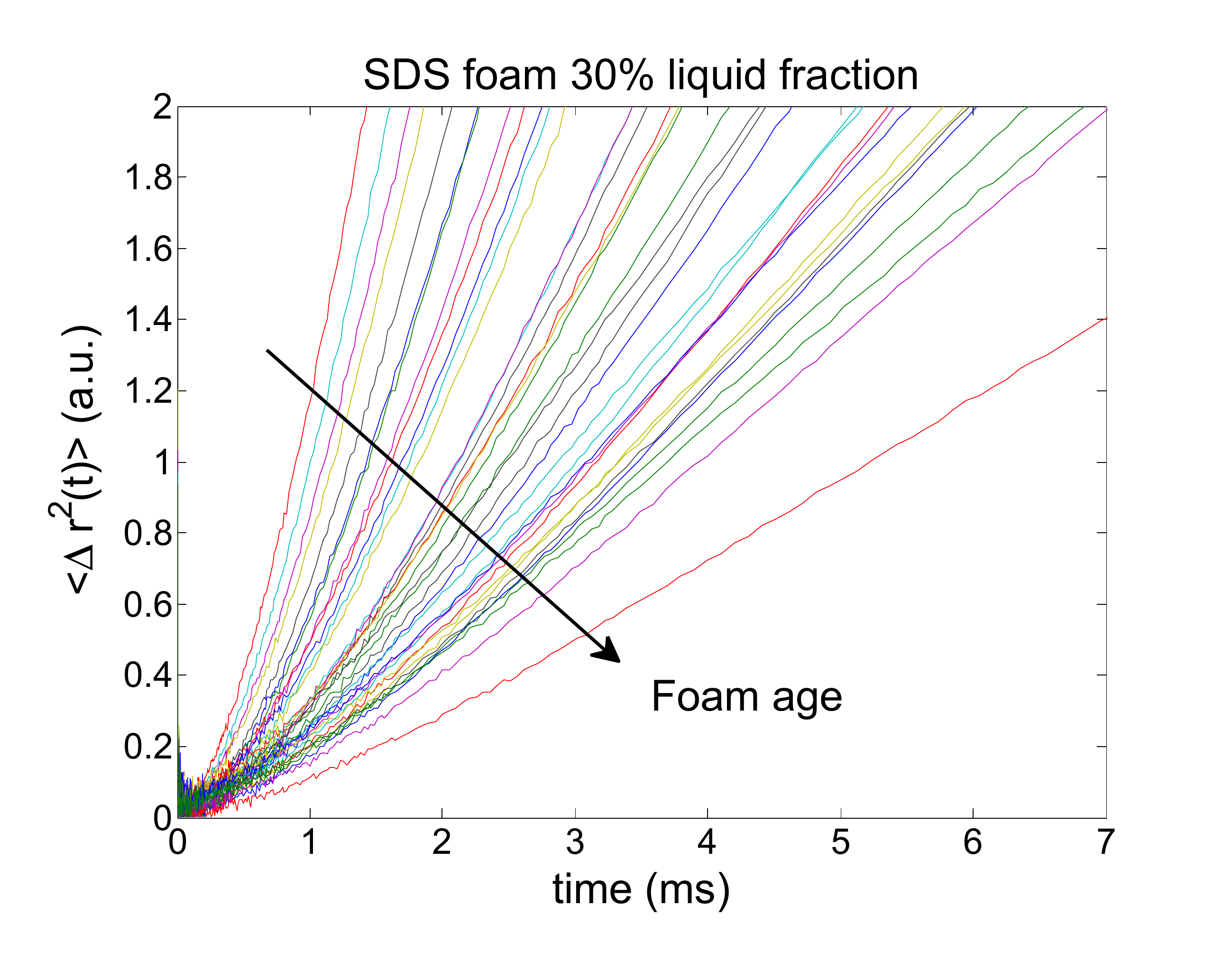}
    \includegraphics[width=0.83\linewidth]{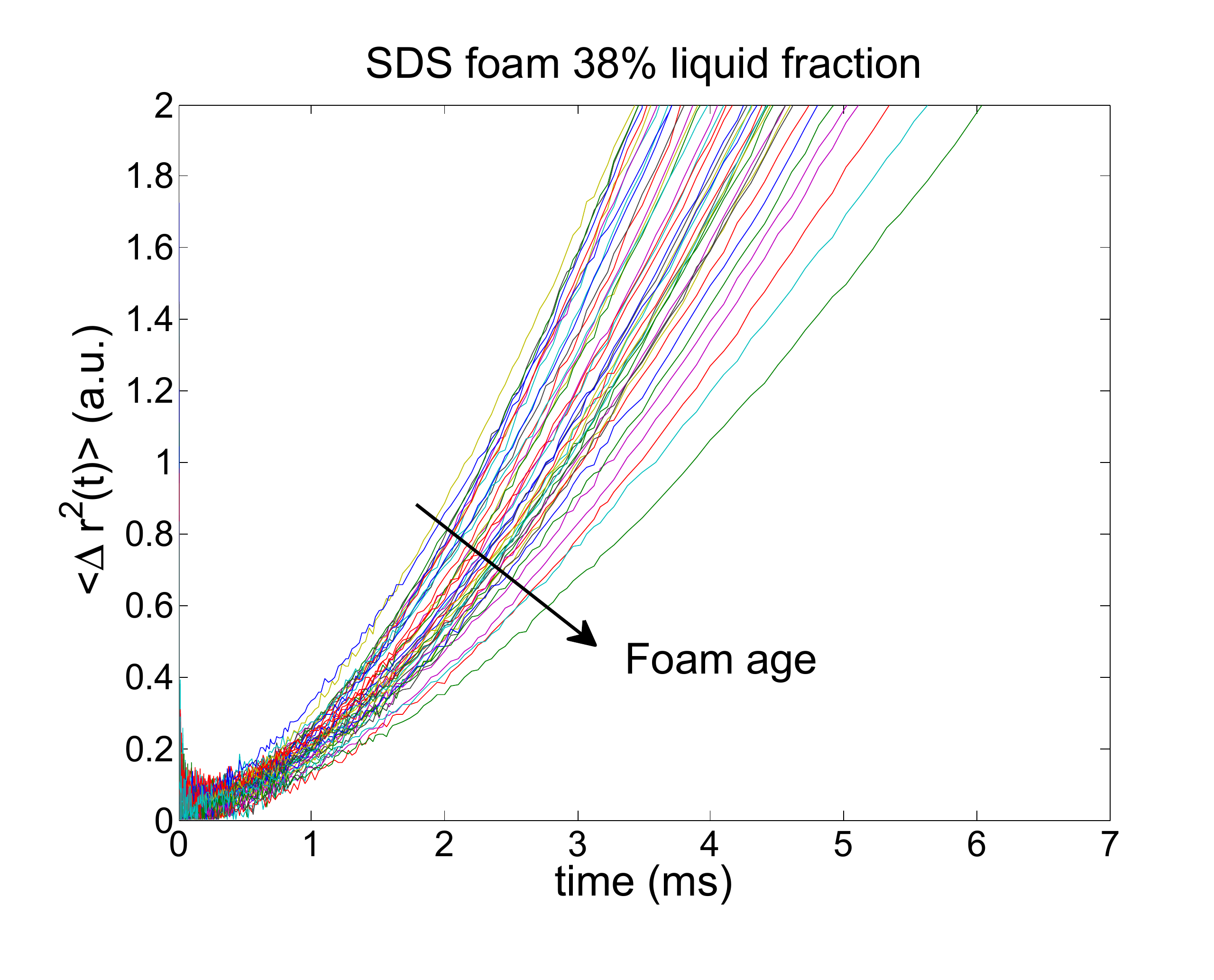}
    \caption{The local dynamics of bubbles as studied by DWS for foams
    of different liquid fraction. a) shows a dry foam (22$\%$ lf), where
    the dynamics is diffusive as can be seen by the linear time evolution
    of the mean squared fluctuation of the scattering interfaces. The length scale
    corresponding to these fluctuations depends on the bubble size,
    thus the diffusivity of the scatterers decreases strongly with age
    of the foam. c) shows a wet foam above the transition (lf 38$\%$).
    Here the increase is quadratic as corresponds to ballistic motion of
    the bubbles, presumably in convective flows induced by the heating.
    The dynamics of these flows does not depend much on the bubble size. b)
    Dynamics in the transition region (lf 30$\%$). Here the dynamics of
    young foams is kinetic as the bubbles are separated at the beginning,
    however as the mean bubble size grows, their dynamics change to a more
    diffusive behavior, which also depends on the age of the foam.   \label{corr} }
\end{figure}

Having observed this transition, it is interesting to investigate
the difference in the local dynamics in the two different regimes
more closely. For this purpose, we determine the mean square
fluctuations of the scattering foam interfaces using DWS
\cite{DWS2}. From the time auto-correlation function $g_2(t)$ of
the transmitted intensity and the transport mean free path, we
obtain $\log(g_2(t)-1)\cdot T^2 \propto \langle \Delta r^2(t)
\rangle$ on short time scales corresponding to the correlation
time \cite{DWS2}. This is shown for foams of different wetness
(i.e. initial liquid fraction) of 22$\%$, 30$\%$ and 38$\%$,  in
Fig. \ref{corr}. The different $\langle \Delta r^2(t) \rangle$
shown for a single foam correspond to different ages of the foam,
where each curve corresponds to a single data point in
Fig.\ref{lstar}. The evolution of aging time is indicated by the
arrow and the ages for the different foams range from 8 to 303 min
for 22$\%$ liquid fraction, 5 to 424 min for 30$\%$ liquid
fraction and 9 to 755 min for 38$\%$ liquid fraction respectively.
As can be seen, for dry foams, $\langle \Delta r^2(t) \rangle$
increases linearly with correlation time indicating a diffusive
dynamics of the scatterers. The slope of the increase directly
indicates the mobility of the scatterers, which can be seen to
decrease with the age of the foam. This decrease corresponds to
the increase in the average bubble size, in the same way as we
have already seen from the mean free path above. For wet foams in
contrast , the dynamics is rather independent of the age of the
foam and the mean square fluctuations increase quadratically with
time. This corresponds to a kinetic dynamics, where the scatterers
travel ballistically during the observed time. Moreover, the time
scale of this kinetic dynamics is independent of the size of the
bubbles, indicating that the bubbles move according to a
convective flow.

In conclusion, using diamagnetically levitated foams not affected
by drainage, we have shown the existence of a transition in the
bulk coarsening dynamics of three dimensional foams at an
effective liquid fraction of 30 $\%$. At lower liquid fraction,
the coarsening dynamics is governed by a von Neumann law
\cite{neumann,generalneumann}, which corresponds to a growth of
the average bubble size with the square root of time. At higher
liquid content, the bubbles grow via Ostwald ripening
\cite{ostwald}, i.e. the average bubble size grows with time to
the power of 1/3. The difference in coarsening dynamics is driven
by the fact that at low liquid content the bubbles are in close
contact thus changing the nature of gas transport between bubbles.

In addition, we have shown that for wet foams, the local dynamics
is kinetic, i.e. the bubbles essentially follow the convective
flow of the interstitial liquid. The time scale of this dynamic
correspondingly is independent of the age of the foam. Thus the
DWS correlation time solely reflects the increase of the mean free
path of light due to the increase in size.

For dry foams in contrast, where bubbles are closely packed, the
dynamics of the bubbles is diffusive as would be expected in a
dense system with many scattering centers. These movements are for
instance induced by local events where bubbles move to relieve
accumulated stresses. As the number of bubbles taking part in such
rearrangements is constant \cite{durian3}, the size evolution of
the bubbles can then also be obtained from the dynamics of these
rearrangements. The time scale of these rearrangements also
increases with the age of the foam as bigger bubbles move more
slowly. In the future it will be interesting to characterize the
dynamics of these rearrangements, which can give a complete
picture of how local dynamics influences the global dynamics of
the foam.

This work was funded by DFG in the context of the IRTG on Soft matter Physics as well as the Landesstiftung
Baden-W\"urttemberg and the Swiss National Science Foundation.

%\color[rgb]{0,0,0}\input acknowledgement.tex   % input acknowledgement

\end{document}